\documentclass[12pt,preprint]{aastex}
\shorttitle{Dust Depletion in two 2175-{\AA} absorbers}
\shortauthors{Jiang et al.}

\begin{document}
\title{High Dust Depletion in Two Intervening Quasar Absorption Line Systems with the 2175-{\AA} Extinction Bump at $z\sim1.4$\footnote{The data presented herein were obtained at the W.M. Keck Observatory, which is operated as a scientific partnership among the California Institute of Technology, the University of California and the National Aeronautics and Space Administration. The Observatory was made possible by the generous financial support of the W.M. Keck Foundation.}}
\author{Peng Jiang\altaffilmark{1,2}, Jian Ge\altaffilmark{2},
J. Xavier Prochaska\altaffilmark{3,4}, Junfeng Wang\altaffilmark{5},
Hongyan Zhou\altaffilmark{1}, and Tinggui Wang\altaffilmark{1}}
\altaffiltext{1}{Key Laboratory for Research in Galaxies and Cosmology, The University of Science and Technology of China, Chinese Academy of Sciences, Hefei, Anhui, 230026, China}
\altaffiltext{2}{Astronomy Department, University of Florida, 211 Bryant Space
Science Center, P. O. Box 112055, Gainesville, FL 32611}
\altaffiltext{3}{University of California Observatories-Lick Observatory, University
of California, Santa Cruz, CA 95064}
\altaffiltext{4}{Department of Astronomy and Astrophysics, University of California,
Santa Cruz, CA 95064}
\altaffiltext{5}{Harvard-Smithsonian Center for Astrophysics, Cambridge, MA 02138}
\email{jpaty@mail.ustc.edu.cn}

\begin{abstract}
We present the column densities of heavy-elements and dust depletion studies in two
strong Mg~II absorption systems at $z\sim1.4$
displaying the 2175-{\AA} dust extinction feature. Column
densities are measured from low-ionization absorption lines using Apparent Optical Depth
Method on the Keck/ESI spectra. We find the dust depletion patterns 
resemble to that of cold diffuse clouds in the Milky Way (MW). The values, [Fe/Zn]$\approx -1.5$
and [Si/Zn]$<-0.67$, are among the highest dust depletion measured for quasar absorption
line systems. In another 2175-{\AA} absorber at $z$=1.64 toward the quasar
SDSS J160457.50+220300.5, Noterdaeme et al. (2009) reported a similar dust depletion
measurement ([Fe/Zn]=$-1.47$ and [Si/Zn]=$-1.07$) and detected C~I and CO absorption lines
on its VLT/UVES spectrum.
We conclude that heavy dust depletion (i.e. a characteristic of cold dense
clouds in MW) is required to produce a pronounced 2175-{\AA} extinction bump.

\end{abstract}


\section{Introduction}
Interstellar dust grains play an important role in the evolution of
galaxies , star formation and planet formation. The size
distribution and composition of dust grains are mainly inferred from
the observed extinction curve and the re-radiated infrared emission.
So far, the best studied interstellar extinction curves are the
ones for the MW (Savage \& Mathis 1979), the Large and
Small Magellanic Clouds (LMC and SMC; Fitzpatrick 1989). The most
conspicuous difference between them is the sequential increment in the
strength of the 2175-{\AA} extinction bump (Stecher 1965)
from SMC to LMC to MW.
The feature is rarely seen in other local galaxies (Keel et al. 2001).

Wang et al. (2004) composed a comprehensive review
on the detection of 2175-{\AA} absorption bump in quasar absorption line (QAL)
systems (see also references therein) and reported
three bumps at high redshift $z\sim 1.4$.
The 2175-{\AA} absorber in several cases (Cohen et al. 1999; Wucknitz et al. 2003)
were confirmed to be Damped Ly$\alpha$ Absorbers (DLAs).
Recently, several new detections have been reported.
Srianand et al. (2008) found a 2175-{\AA} extinction feature in two Mg~II
systems at z$\sim$1.3 and detected associated 21-cm absorption, which usually
traces cold gas content. Noterdaeme et al. (2009)
presented a detection of carbon monoxide molecules (CO) at $z$ = 1.6408
toward a red quasar and a pronounced 2175-{\AA} bump at the same redshift.
A super-strong 2175-{\AA} absorption galaxy at $z$=0.8839 toward the quasar
SDSS J100713.68+285348.4 was reported by Zhou et al. (2010).
Besides the quasar absorption approach to search for the 2175-{\AA} dust extinction
feature in high redshift galaxies, 
the analysis of (Gamma-Ray Burst) GRB afterglow spectra has also
revealed several positive detections from intervening absorbers and from
gas in the GRB host galaxy (e.g., Ellison et al. 2006; El{\'i}asd{\'o}ttir et al. 2009;
Prochaska et al. 2009).

The physical conditions of absorbers, such as ionization state,
electron density, temperature, gas kinetics, metallicity and dust depletion, can
be measured from the relative strength and the profile of metal absorption lines
(e.g. Pettini et al. 1994; Prochaska \& Wolfe 1998;
Jenkins \& Tripp 2001; Prochaska et al. 2002). The measurements of metal
absorption lines associated with very rare 2175-{\AA} quasar absorbers
can definitely help us to understand the physical characteristics giving rise to the
dust extinction features at high redshift (i.e. $z>1.0$).
In this paper, we analyze the dust depletion levels of
the two 2175-{\AA} absorbers in lines of sight toward the quasars
SDSS\footnote{Sloan Digital Sky Survey; http://www.sdss.org/}
J012147.73+002718.7 (hereafter J0121+0027; $z$=2.2241)
and SDSS J145907.19+002401.2 (hereafter J1459+0024; $z$=3.0124).

Dust depletion in MW have been widely studied over
$\sim$240 sight lines in the past several decades with the
ultraviolet (UV) space telescopes (e.g. Jenkins 2009 and references therein).
Significant deviation of depletion levels in different Galactic
environments is found. Generally speaking, heavy elements are more depleted onto
dust grains in cold gas clouds than in warm and halo gas clouds
(e.g. Sembach \& Savage 1996; Welty et al. 1997; Welty et al. 1999).
Comparing the dust depletion patterns of the 2175-{\AA} absorbers with
the Galactic clouds can provide important clues on the general astrophysical
characteristics of them.
The dust depletion for the two absorbers in this work
is very similar to that measured in cold clouds of MW, which are characterized
by cold, dense and primarily molecular gas.
Dust depletion is also usually used to identify the dust content in DLAs
(e.g. Pettini et al. 1994, 1997, 1999; Lu et al. 1996).

This paper is organized as follows. In \S2 we describe the observations;
\S3 presents the analysis of UV extinction bumps;
in \S4 we present the measurements of elemental column densities
in the two absorption line systems; in \S5 we explore the dust depletion patterns;
discussions are presented in \S6; future work and conclusions are given in \S7.

\section{Observations}
The SDSS spectra of these two quasars
are remarkable with depressed flux around $\sim 5500$ \AA. They were
initially uncovered by Patrick B. Hall, when he carried out a visual
inspection of the SDSS spectra of the DR1 quasars with $z\ge1.6$
(Schneider et al. 2003) to identify unusual quasar subtypes
(i.e. primarily broad absorption line (BAL) quasars, nitrogen-strong
quasars and dust reddened quasars).
The features were then interpreted as 2175-{\AA} extinction bumps associated with
two strong intervening Mg~II absorbers at $z\sim$1.4 in lines of sight toward
the quasars by Wang et al. (2004).\footnote{The SDSS spectra in
Wang et al. (2004) were extracted from Data Release 1 (DR1; Abazajian et al. 2003),
while the SDSS spectra in this work are extracted from DR7 (Abazajian et al. 2009).
All the SDSS spectra in this work are corrected for the Galactic reddening by
using the dust map of Schlegel et al. (1998) before being analyzed.}
The point-spread function magnitudes measured from the SDSS
images are $u=20.68\pm0.06$, $g=19.09\pm0.01$, $r=18.46\pm0.01$,
$i=17.93\pm0.01$, and $z=17.65\pm0.02$ for J1459+0024;
$u=20.44\pm0.06$, $g=19.97\pm0.02$, $r=19.38\pm0.01$, $i=19.03\pm0.02$
and $z=18.70\pm0.04$ for J0121+0027,
respectively. The SDSS spectra cover $\sim 3800-9200$ \AA~with a spectral
resolution $R \sim2000$ (Stoughton et al. 2002).

We performed follow-up spectroscopic observations of the quasars at higher
spectral resolution to explore the dust depletion in these two 2175-{\AA} absorbers.
The medium-resolution ($R\sim10000$), medium signal-to-noise ratio ($SNR\ge10$)
quasar spectra were obtained with the 10-m Keck telescope using the Echellette Spectrographs
and Imager (ESI; Sheinis et al. 2002). The entire spectra cover the full optical range from 
3900 \AA\ to 1.1 $\mu$m, recorded by a single 2K$\times$4K Lincoln Labs CCD with 15 $\mu$m
pixels. The quasars were observed on 20 December 2003 (UT) with total exposure times of
900s and 2700s for J1459+0024 and J0121+0027 under good conditions but variable seeing of
approximately FWHM=$0.8''$.
We implemented the 0.5$''$ slit for J1459+0024 and the 0.75$''$ slit for
J0121+0027 which yields a FWHM resolution of 33 km/s and 44 km/s respectively.
An additional 1800s exposure for J1459+0024 was obtained on 18
February 2004 (UT). The data were
reduced and calibrated using the ESIRedux software package (v1.0) developed by JXP
(see Prochaska et al. 2003 for details).
Then the two reduced spectra of J1459+0024 were combined. The data
were normalized by fitting a series of polynomials to absorption-free
regions of the quasar spectrum.
The metal absorption lines of interest are plotted in velocity space in Figures 5 and 6.

\section{2175-{\AA} Extinction Bumps}
Jiang et al. (2010) used a parameterized extinction curve (FM parameterization;
Fitzpatrick \& Massa 1990) constituted by
a linear component and a Drude component to describe the optical/UV extinction
curve in the rest frame of quasar absorber.
The linear component is used to model the underlying extinction\footnote{The linear
component also accounts for the variation of the intrinsic quasar spectral slopes.},
while the Drude component is used to model
the possible 2175-{\AA} extinction bump. The parameterized extinction curve is
written as
\begin{equation}
A(\lambda)=c_1+c_2x+c_3D(x,x_0,\gamma)
\end{equation}
where $x=\lambda^{-1}$.
And $D(x,x_0,\gamma)$ is a Drude profile, which is expressed as
\begin{equation}
D(x,x_0,\gamma)={\frac{x^2}{(x^2-x_0^2)^2+x^2\gamma^2}}
\end{equation}
where $x_0$ and $\gamma$ is the peak position and FWHM
of Drude profile, respectively. Our aim is to unveil the 2175-{\AA}
absorption feature associated with absorption line systems on quasar spectra.
We do not try to derive the absolute extinction curve. Our derived curve
is a relative extinction curve without being normalized by E(\bv). We cannot measure the
conventional extinction parameters $A_V$, E(\bv) and $R_V$ from it.
But all the features of 2175-{\AA} absorption bump are preserved.
The strength of bump is measured
by the area of bump $A_{bump}=\pi c_3/(2\gamma)$, which can be interpreted
as rescaling the integrated apparent optical depth of bump absorption
($A_{\lambda}={\frac{2.5}{ln 10}}\tau_{\lambda}$). 
We fit the SDSS spectra of J1459+0024 and J0121+0027 by reddening
the SDSS composite quasar spectrum (Vanden Berk et al. 2001) with parameterized
extinction curves at the redshifts $z_{abs}=1.3947$ for the former and $z_{abs}=1.3888$
for the later, respectively.
The redshifts are those of the strongest component in the absorption line profiles,
which are measured on our Keck/ESI spectra, for the two 2175-{\AA} absorbers.
Since the 2175-{\AA} extinction bump is a very broad feature,
whose FWHM is usually greater than 300{\AA} (Fitzpatrick \& Massa 2007), the small differences
between these redshifts and the redshifts of the centroid of absorption line profiles could
not affect the results of spectrum fitting in this work.
To focus on fitting the continuum of quasar spectrum, the regions with strong emission lines
and known strong absorption lines are masked.
Several intervening Ly$\alpha$ absorption lines are present blueward of the Ly$\alpha$ emission lines
on the spectrum of J1459+0024 with $z>$2.6. We remove them iteratively by clipping the outliers
beyond 4$\sigma$ error. The results of spectrum fitting are listed
in Table 1. The best models are overplotted with SDSS data in Figures 1a and 2a. To emphasize
the requirement of a absorption bump on the extinction curve, we also overplot the
reddened composite quasar spectrum by using the linear component of best model
(green lines in Figures 1a and 2a) only. The required bump of J1459+0024 is so broad that
its wings exceed the wavelength coverage of SDSS spectrum. Thus, the entire observed spectrum
of J1459+0024 is completely off from the green model in Figure 1a.

The intrinsic variation of quasar spectra can mimic an
extinction bump in some cases (Pitman et al. 2000). We gauge the significance of
the extinction bumps in this work with the simulation technique developed by
Jiang et al. (2010). The simulation technique begins with the selection of a
control sample of SDSS quasar spectra at a similar redshift to the quasar of interest.
Then we fit each of them by reddening the composite quasar spectrum
with a parameterized extinction curve at redshift of the absorber of interest.
The parameters $x_0$ and $\gamma$ in the parameterized extinction curve are fixed
to the best values fitting the 2175-{\AA} extinction bump of interest.
The distribution of bump strengths is expected to be a Gaussian
by assuming random fluctuation of continuum on each spectrum in the control sample.
If the bump strength of the absorber is far away from
this distribution, then the bump has statistical significance.
For each quasar, we select the spectra in SDSS DR7 database classified as quasars
({\tt specClass=QSO} or {\tt HIZ\_QSO}) with redshift
in the range of $z_{emi}-0.05 < z < z_{emi}+0.05$ and signal-to-noise ratio
{\tt SN\_I} $\ge$ 6\footnote{Most of the false positive detections of quasar can
be rejected by this criterion. {\tt SN\_I}=18 for J1459+0024 and {\tt SN\_I}
=8 for J0121+0027.}, as its control sample.
The resultant control sample of J1459+0024 is constituted by 1002 SDSS quasars around $z=3.0124$.
The resultant control sample of J0121+0027 is constituted by 2351 SDSS quasars around $z=2.2241$.
The extracted distributions of bump strengths are presented in Figure 1c and 2c.
Both of the bumps in J1459+0024 and J0121+0027 are statistically significant at
confidence level of $>5\sigma$. The width of bump strength distribution
for J1459+0024 is much greater than that for J0121+0027. That is because the bump on
the spectrum of J1459+0024 ($\gamma=2.68 {\mu}$m$^{-1}$)
is much broader than that of J0121+0027 ($\gamma=0.80 {\mu}$m$^{-1}$).
For the spectrum without a real extinction feature, the fitted bump strength is
an integration of random fluctuations over the wavelength range covered by
the bump of interest. 
In principle, a longer integration makes the resultant distribution
of bump strengths broader.
Therefore, a broad bump has to be stronger (with larger bump area)
than a narrow one in order to be significantly
distinguished from random fluctuations of quasar spectra.
In theory, the mean value of bump strength distribution should be zero.
But, the mean value of the bump strength distribution is about 1.0 for
J1459+0024 (see Figure 1c). The significant non-zero mean bump strength is mainly caused
by the low quality of the SDSS quasar composite spectrum in far UV band
of its rest frame.
There are only about 150 low (or moderate) observed quasar spectra being used
to create the composite spectrum in far UV band (Vanden Berk et al. 2001)
\footnote{There are 2204 quasar spectra being used to create the SDSS quasar
composite spectrum in total.}.
Thus, it is very likely that the composite quasar spectrum cannot represent
the average properties of quasar spectra in far UV band completely.

We also perform continuum fitting by reddening the SDSS composite spectrum
with different types of average extinction curves, namely that from
the SMC, the LMC Supershell (LMC2), the LMC (Gordon et al. 2003) and
the MW (Fitzpatrick \& Massa 2007).
The 2175-{\AA} extinction bump is absent in the SMC extinction curve and it is strongest in the
LMC and the MW extinction curves. The strength of bump in LMC2 extinction curve is medium.
These extinction curves are given in the FM parameterization.
The two free parameters in the continuum fitting are E(\bv) and a normalization scale. The
results of spectrum fitting are presented in Table 2. We show the models using different
extinction curves for the two reddened quasar spectra in Figure 3.
The spectrum of J0121+0027 can be fitted by the average LMC
($\chi_{\nu}^2 = 1.14$) and the average MW ($\chi_{\nu}^2 = 1.10$) extinction curve models.
But the spectrum of J1459+0024 cannot be fitted by any model ($\chi_{\nu}^2 = 3.88$ in the
best case). That is because the extinction bump on J1459+0027 is much broader
($\gamma=2.68\pm 0.07 {\mu}$m$^{-1}$) than that of the average
LMC ($\gamma=0.934\pm0.016 {\mu}$m$^{-1}$) extinction curve or
the average MW ($\gamma=0.922 {\mu}$m$^{-1}$) extinction curve.

We compare the two extracted 2175-{\AA} absorption bumps at $z\sim$1.4 with the
bumps observed in the MW (see Figure 4; Fitzpatrick \& Massa (2007)).
The peak position and strength of bumps are
within the range of Galactic values. However, the FWHM of bump in J1459+0024 is
larger than the broadest Galactic bump, which is measured in the line of sight toward the star
HD29647. The direct comparison of Optical/UV extinction curves of J1459+0024
and HD29647 is plotted in Figure 3c and the comparison of
J0121+0027 and HD164816 (the latter has a similar bump with the former)
is in Figure 3d. Since the extracted extinction curves for quasar absorbers
are relative extinction curves, we cannot measure the absolute visual extinction
$A_V$ on them. We add an arbitrary normalization on the extracted curves
while plotting them with Galactic curves.

\section{Measurements of Column Densities}
The total number of absorption components are not fully resolved because of our limited
resolution and the intrinsic line-blending. For the line of sight toward J0121+0027,
three Mg~II absorption systems are seen at $z=$1.3877, 1.3888, 1.3907, which
match the approximate redshift of 2175-{\AA} absorption bump
($z\sim$1.39; Wang et al 2004) detected in the SDSS spectrum.
Since the first two strong systems are only separated by $\sim$ 140 km s$^{-1}$
(corresponding to $z$=1.3888), they are
considered as two associated components in one absorption system at $z$=1.3888 in this work.
The third one could be another component,but it is neglected in our analysis because it is
too weak to affect the measurements.
For the line of sight toward J1459+0024, one strong Mg~II absorption system is seen at
$z$=1.3947. We identified numerous absorption lines associated with the
absorption system, consistent with the 2175-{\AA} absorption
bump feature redshift ($z\sim$1.39) measured in the SDSS spectrum.

We carry out the Apparent Optical Depth Method (AODM; Savage \& Sembach 1991) to measure
the column densities of low ionization absorption lines in those two 2175-{\AA} absorption
systems. 
In general, the column densities are measured by integrating the optical depth
over the velocity range of the spectra covering all of the
absorption in the detected transitions. The uncertainties for detections are reported
as 1$\sigma$ errors. The upper limits are measured by integrating the optical depth
over the same velocity range as an unsaturated line (i.e. Fe~II$\lambda$2374) and are reported
as 3$\sigma$ statistical limits. For the saturated lines, the column densities derived
from the AODM are reported as lower limits.
We also measured the column densities of transitions in red and blue components
of the absorption system toward J0121+0027 separately.
The boundary of these two components is chosen at the point of $-70$ km s$^{-1}$ compared
to redshift $z$=1.3888 (the thin blue dot line in Figure 6). All ionic
column densities are presented in Table 3.

\section{Dust Depletion Patterns}
The column density measurements of ions from absorption lines only
represent the elements in the gas phase. If one assumes that
ionization corrections are small for what is expected to be a
primarily neutral gas, the observed relative gas-phase abundances
reflect the underlying nucleosynthetic abundance pattern modified
by the differential depletion of the elements.
Frequently, the relative abundance,
[Fe/Zn]=log(N(Fe)/N(Zn))$-$log(N(Fe)/N(Zn))$_{\sun}$\footnote{The possible
different underlying nucleosynthetic pattern can induce an error at
the level of a few tenths dex (e.g. Lu et al. 1996; Prochaska et al. 2000).
This error is likely to decrease in high metallicity absorbers, where 
less massive stars are more likely to dominate nucleosynthetic procedure.},
is used to infer the dust depletion level, even when N(H~I) is not obtainable,
by assuming that Zn, which is assumed to be nearly undepleted due to
its low condensation temperature (Meyer \& Roth 1990; Pettini et al. 1994),
tracks the Fe Peak closely in nucleosynthesis.

The detected transitions of zinc are
Zn~II$\lambda$2026 and Zn~II$\lambda$2062 on our Keck spectra.
These two absorption lines are weakly blended with Mg~I$\lambda$2026
and Cr~II$\lambda$2062 lines, respectively. Since the oscillator strength of transition
Cr~II$\lambda$2062 is fairly weak and other transitions of Cr~II are only marginally detected
in both of the absorption systems, we think the absorption line of
Zn~II$\lambda$2062 is not contaminated by blending
and take the column density of zinc measured on it
to derive dust depletion patterns. For other species, the weighted mean values are adopted if
there are two or more transitions detected.
The depletion patterns are listed in Table 4.

We compare the dust depletion patterns in the two 2175-{\AA} absorption systems with
the depletion patterns observed in SMC ISM, LMC ISM and cold Galactic
disk clouds (Welty et al. 1997; Welty et al. 1999) in Figure 7. The dust depletion
levels increase sequentially from SMC, LMC to MW. It is clear
that the depletion patterns of both absorption systems closely resemble that in
cold Galactic clouds, especially from the best measured [Fe/Zn] values.
The strength of 2175-{\AA} extinction bumps increases in the same manner of
dust depletion from SMC, LMC to MW. It is very likely that heavy dust depletion
is required to give rise of a 2175-{\AA} absorption. See more discussions in
\S6.

The depletion levels, [Fe/Zn]$\approx -1.5$ and [Si/Zn]$<-0.67$, in both of absorbers
are among the highest dust depletion ever measured for QAL systems.
The mean value of [Fe/Zn]$>$-0.6 was measured by York et al. (2006) with a large
sample constituted by 809 Mg~II absorption systems at $1\le z<2$.
A comparable depletion measurement with ours has been reported by
Petitjean et al. (2002) in a DLA bearing molecular hydrogen
with [Fe/Zn]=-1.59 at $z$=1.973 toward the quasar Q0013-004.
In addition, Noterdaeme et al. (2009) measured a similar high dust depletion
level ([Fe/Zn]=-1.47 and [Si/Zn]=-1.07) of a 2175-{\AA} absorber at $z$=1.64 toward the
quasar SDSS J160457.50+220300.5 with VLT/UVES.
We compare the dust depletion level in this work with a combined DLA/sub-DLA
sample, for which high resolution spectroscopic data is available
(Keck/ESI and Keck/HIRES data from Prochaska et al. 2007
; VLT/UVES data from Noterdaeme et al. 2008), in Figure 8. We ignore
censored data in the sample when plotting the figure.
To present the dust depletion of two 2175-{\AA} absorbers in the same figure,
we assume the column density of neutral hydrogen
N(H~I)=10$^{21}$ cm$^{-1}$ in both of them.
This assumption is arbitrary and is for drawing purpose only.
DLAs bearing H$_2$ are marked with black filled circles.
It is clear that DLAs showing high dust depletion level,
[Fe/X]\footnote{X could be Zn, Si and S.}$<$-0.4, have a high H$_2$ detection rate (e.g.,
Ge \& Bechtold 1997; Ge et al. 2001; Ledoux et al. 2003; Cui et al. 2005; Noterdaeme et al. 2008).
It is reasonable to assume, therefore, that the gas in these two 2175-{\AA} absorbers
bears molecules.

\section{Discussion}
The dust depletion patterns ([X/Zn]) in the SMC ISM is very similar to that measured in Galactic halo
clouds, while the depletion in the LMC ISM is very similar to that in Galactic warm disk clouds
(see Table 4).
Since the overall metallicity and dust-to-gas ratio are very different in SMC,
LMC and MW (e.g. Pei 1992; Welty et al. 1997), the similarities
probably indicate that dust depletion can hardly be dependent
on these two physical parameters. In the MW, the most severe depletions are found in lines of
sight with the largest mean densities and largest fraction of H$_2$;
the least severe depletions are observed for halo clouds and for high velocity clouds
in the Galactic disk (e.g. Jenkins, Wallerstein and Silk
1984; Savage \& Sembach 1996; Fitzpatrick 1996; Trapero et al. 1996;
Jenkins \& Wallerstein 1996; Welty et al. 1997).
Such a dependence of dust depletion on environment probably has to do with the
modification or destruction of dust grains by supernova shocks. In cold dense
molecular clouds, large dust grains can form substantially by coagulation of finer grains.
While in warmer environments, large grains are usually destroyed by energetic processes
and some depleted elements return to the gas phase.

The fact that the strength of 2175-{\AA} extinction bumps becomes weaker with the
decreasing dust depletion levels from MW, LMC to SMC may indicate the destruction
of 2175-{\AA} absorption carriers happens simultaneously with the destruction of large grains.
The large polycyclic aromatic hydrocarbon (PAH) molecules, which have strong
$\pi \rightarrow \pi^*$ absorption in the 2000--2500 {\AA} region, are proposed
to be the carrier of 2175-{\AA} absorption (Li \& Draine 2001). These large molecules
can be easily destroyed by photon-thermo dissociation, coulomb explosion and/or
X-ray destruction (e.g. Voit 1992 and reference therein).

We search for the literatures and find 15 lines of sight in MW
having both measurements of UV absorption bump and metal abundance available (see Table 5).
Dust depletion is measured with the abundance ratio [Fe/H] by assuming the
underlying abundances are solar values.
The measurements of Galactic 2175-{\AA} extinction bump are taken
from Fitzpatrick \& Massa (2007). Note that the area of bump defined
in FM07 is different from that in this work. 
Since the extinction curve has been normalized by E(\bv) in FM07,
$A_{bump}$=E(\bv)$\times A_{bump}^{*}$, where $A_{bump}^{*}$ is the area defined in FM07.
We plot the relative bump strength ($\frac{A_{bump}}{A_V}$) versus [Fe/H] in Figure 9a.
We calculate Spearman's $\rho$, a non-parametric measure of statistical dependence
between two variables, for the relative strength of bump and [Fe/H].
The resultant Spearman's $\rho = -0.46$ suggests a tentative anti-correlation between them
with statistical confidence level of 90\%. The fairly large scatter of bump strength
could be caused by the varying size of PAH molecules from one sightline to another.
Draine (2003) interpreted the observed variations in FWHM (and small variations in
peak position) of Galactic 2175-{\AA} extinction bump profile would result from
differences in the PAH mix. This differences could also affect the oscillator strength
per molecule and lead to the deviation of bump strengths.
Note that the two 2175-{\AA} extinction bumps in this work have very different widths
and strengths.

In Figure 9c, we confirm the anti-correlation
between f(H$_2$) and [Fe/H] (Spearman's $\rho = -0.81$) and interpret it as
simultaneous dissociation of molecular hydrogen with destruction of large dust grain.
In addition, we find a tentative correlation between dust-to-gas ratio and [Fe/H] with
Spearman's $\rho = +0.54$ (Figure 9d). It is consistent with the grain destruction scenario.
Assuming a constant density and a sphere shape of dust grains, the total cross
section of finer grains would be greater than the total cross section of large grains even
if some materials of dust grains are evaporated during the destruction processes.
We summarize the discussions as a molecule and large grain destruction scenario: 1. molecules
(e.g. PAH and H$_2$) and large dust grains form substantially in cool and dense environment;
2. molecules and large grains would be destroyed simultaneously when the
environment is heated (e.g. heated by supernova
shock waves; irradiated by extreme UV photons); 3. the destruction of large dust
grains releases some depleted elements into the gas phase; the dissociation of
H$_2$ molecules decreases the f(H$_2$); the dissociation of PAH molecules diminishes
the strength of 2175-{\AA} absorption bump.

\section{Future Work and Conclusion}
In this work, we measure the dust depletion levels for the two 2175-{\AA} absorbers.
More observations in other wavelength bands will definitely help us to unveil the
astrophysical conditions giving rise to 2175-{\AA} absorption. For the two absorbers
at $z\sim1.4$ in
this work, the HST/COS\footnote{http://www.stsci.edu/hst/cos/} spectrometer covers the
important H~I and H$_2$ transitions and can be used to determine the gas metallicity
and f(H$_2$). We are planning to propose Keck/HIRES observations
to catch the C~I and other transitions blueward of the ESI coverage to the atmospheric
cutoff. The atomic carbon is an excellent tracer of cold, dense gas
because of its low ionization potential (below 1 Ryd). The ratios of C~I and C~II
transitions will provide us a powerful tool to probe the physical environment of
the absorbers.

By using the parameterized extinction curve fitting technique, we are searching
for more 2175-{\AA} absorbers in SDSS quasar spectra database. Our
preliminary result shows 18 detections of 2175-{\AA} extinction bump associated with
dusty strong Mg~II quasar absorbers in SDSS DR3. 

In this paper, we report on follow-up moderate resolution spectroscopy of the
2175-{\AA} absorption systems at $z\sim$1.4 in lines of sight toward two quasars
J0121+0027 and J1459+0024. The column densities of heavy elements are
measured by low ionization absorption lines with AODM. We derived the
dust depletion patterns of both absorption systems and found they closely
resemble to that of cold diffuse disk clouds in MW.
The values, [Fe/Zn]$\approx -1.5$
and [Si/Zn]$<-0.67$, are among the highest dust depletion measured for quasar absorption
line systems. We conclude that heavy dust depletion (i.e. a characteristic of cold dense
clouds in MW) is required to produce a pronounced 2175-{\AA} extinction bump.

\acknowledgements
This work was partially supported by NSF with grant
NSF AST-0451407, AST-0451408 \& AST-0705139
and a China NSF grant (NSF-10973012).
P.J acknowledges support from China Scholarship Council.
This research has also been partially supported by the
CAS/SAFEA International Partnership Program for Creative Research Teams.
J.X.P is partially supported by an NSF CAREER grant (AST--0548180)
and by NSF grant AST-0908910.

The authors wish to recognize and acknowledge the very significant cultural
role and reverence that the summit of Mauna Kea has always had within the
indigenous Hawaiian community.  We are most fortunate to have the opportunity
to conduct observations from this mountain.

Funding for the SDSS and SDSS-II has been provided by the Alfred P. Sloan
Foundation, the Participating Institutions, the National Science Foundation,
the U.S. Department of Energy, the National Aeronautics and Space Administration,
the Japanese Monbukagakusho, the Max Planck Society, and the Higher Education
Funding Council for England. The SDSS Web Site is http://www.sdss.org/.

The SDSS is managed by the Astrophysical Research Consortium for the
Participating Institutions. The Participating Institutions are the American
Museum of Natural History, Astrophysical Institute Potsdam, University of Basel,
University of Cambridge, Case Western Reserve University, University of Chicago,
Drexel University, Fermilab, the Institute for Advanced Study, the Japan
Participation Group, Johns Hopkins University, the Joint Institute for Nuclear
Astrophysics, the Kavli Institute for Particle Astrophysics and Cosmology,
the Korean Scientist Group, the Chinese Academy of Sciences (LAMOST),
Los Alamos National Laboratory, the Max-Planck-Institute for Astronomy (MPIA),
the Max-Planck-Institute for Astrophysics (MPA), New Mexico State University,
Ohio State University, University of Pittsburgh, University of Portsmouth,
Princeton University, the United States Naval Observatory, and the University of Washington.

\clearpage

\begin{figure}
\epsscale{1.0}
\plotone{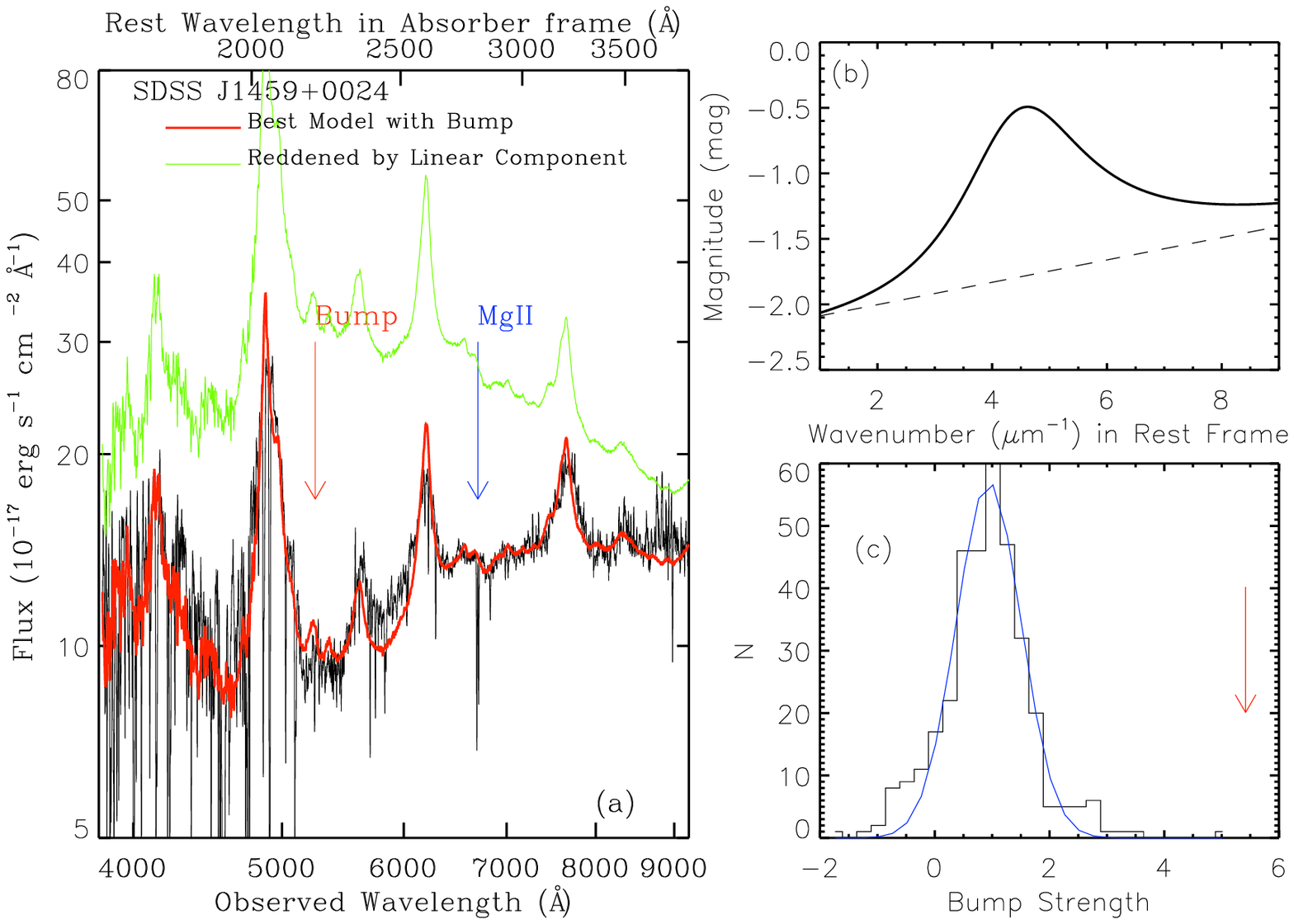}
\caption{The best fitted extinction model for J1459+0024. (a). Red solid line is the best fitted model. Red arrow indicates the center of extinction bump and blue arrow indicates the Mg~II absorption lines. Green solid line is reddened composite quasar spectrum by using the linear component of best model only to emphasize the requirement of extinction bump. (b). The best fitted extinction curve (see the parameters of extinction curves in Table 1). (c). Distribution of fitted bump strength of the control sample for J1459+0024. The blue line is the best fitted Gaussian profile. Red arrow indicates the strength of bump on the spectrum of J1459+0024.\label{fig1}}
\end{figure}
\clearpage

\begin{figure}
\epsscale{1.0}
\plotone{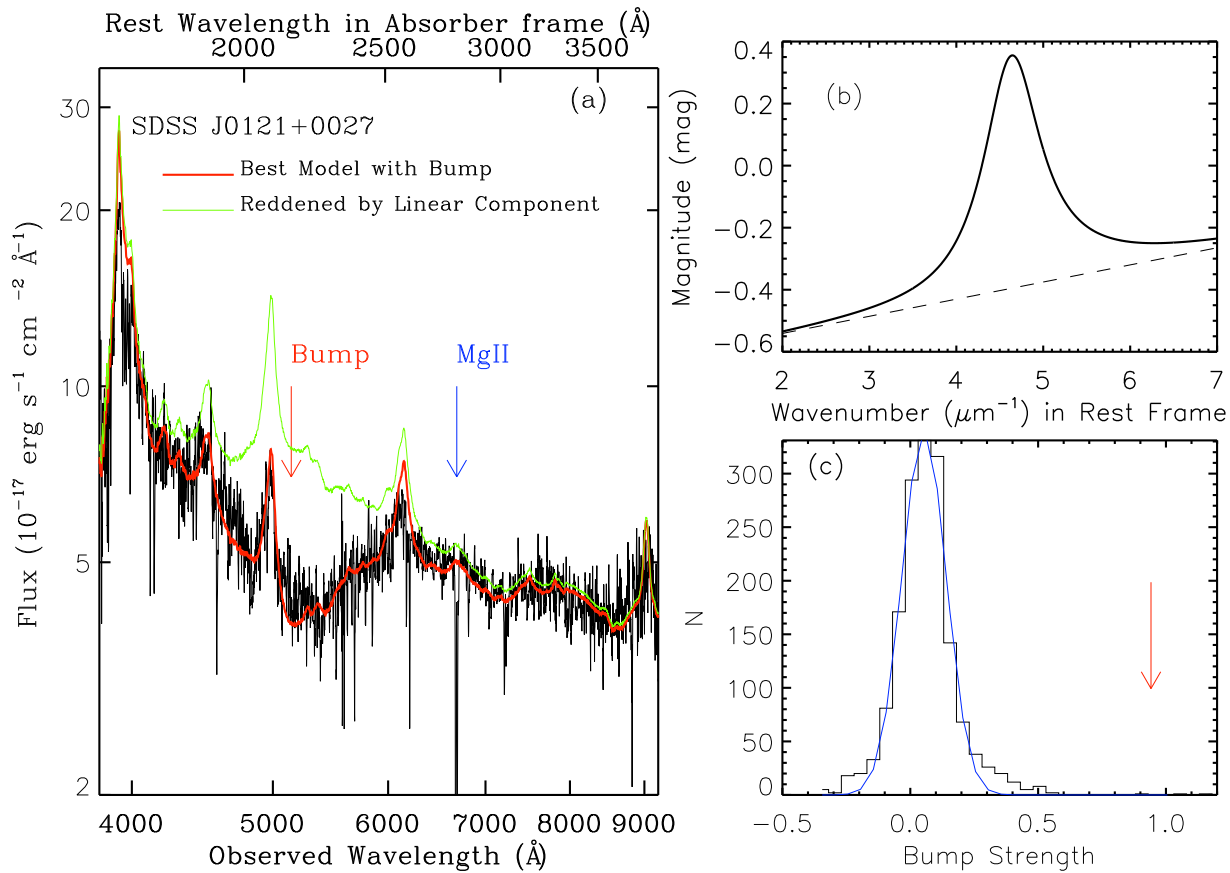}
\caption{The best fitted extinction model for J0121+0027. (a): Red solid line is the best fitted model. Green solid line is reddened composite quasar spectrum by using the linear component of best model only. (b): The best fitted extinction curve. (c): Distribution of fitted bump strength of the control sample for J0121+0027. Red arrow indicates the strength of bump on the spectrum of J0121+0027.\label{fig2}}
\end{figure}

\clearpage
\begin{figure}
\epsscale{1.0}
\plottwo{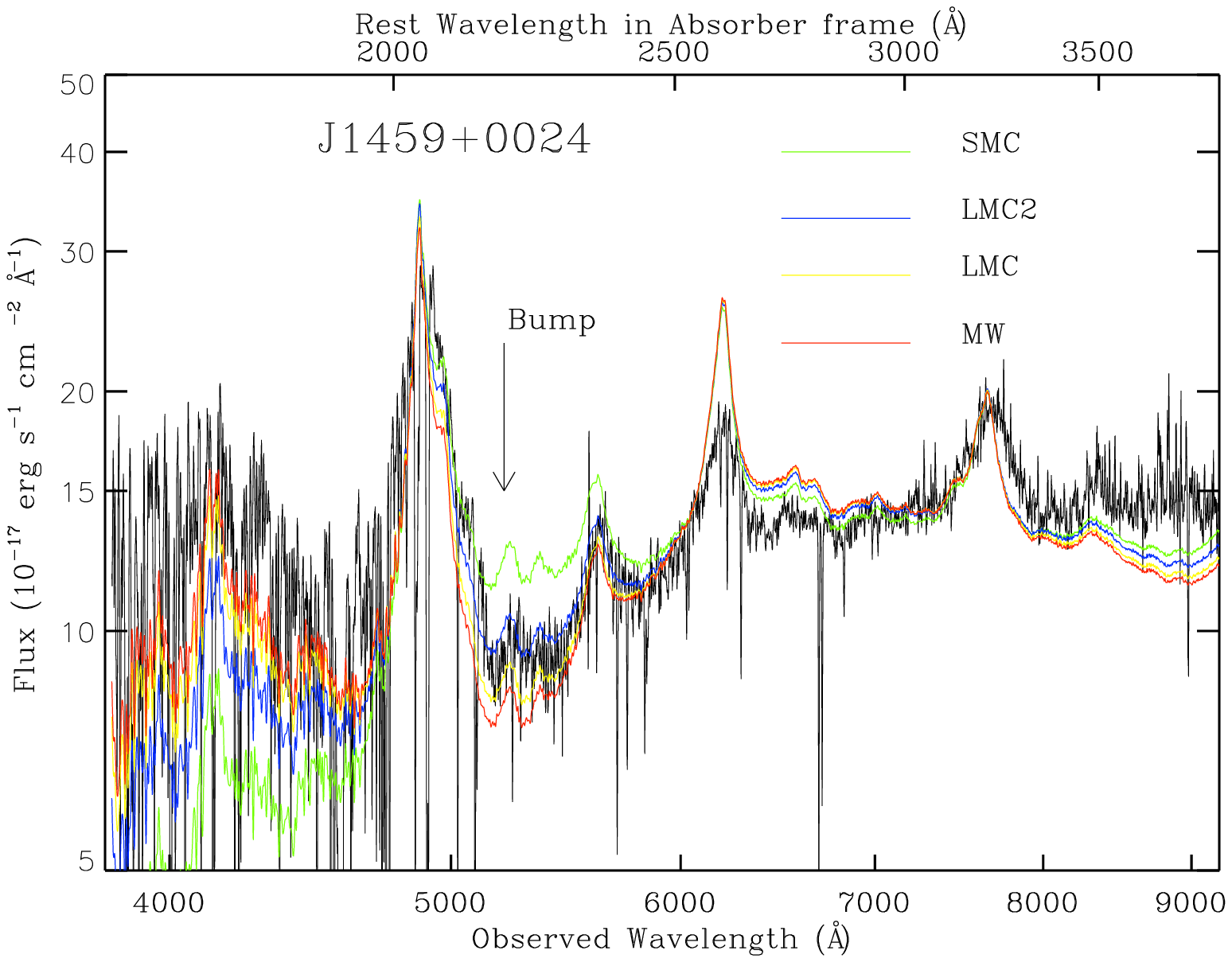}{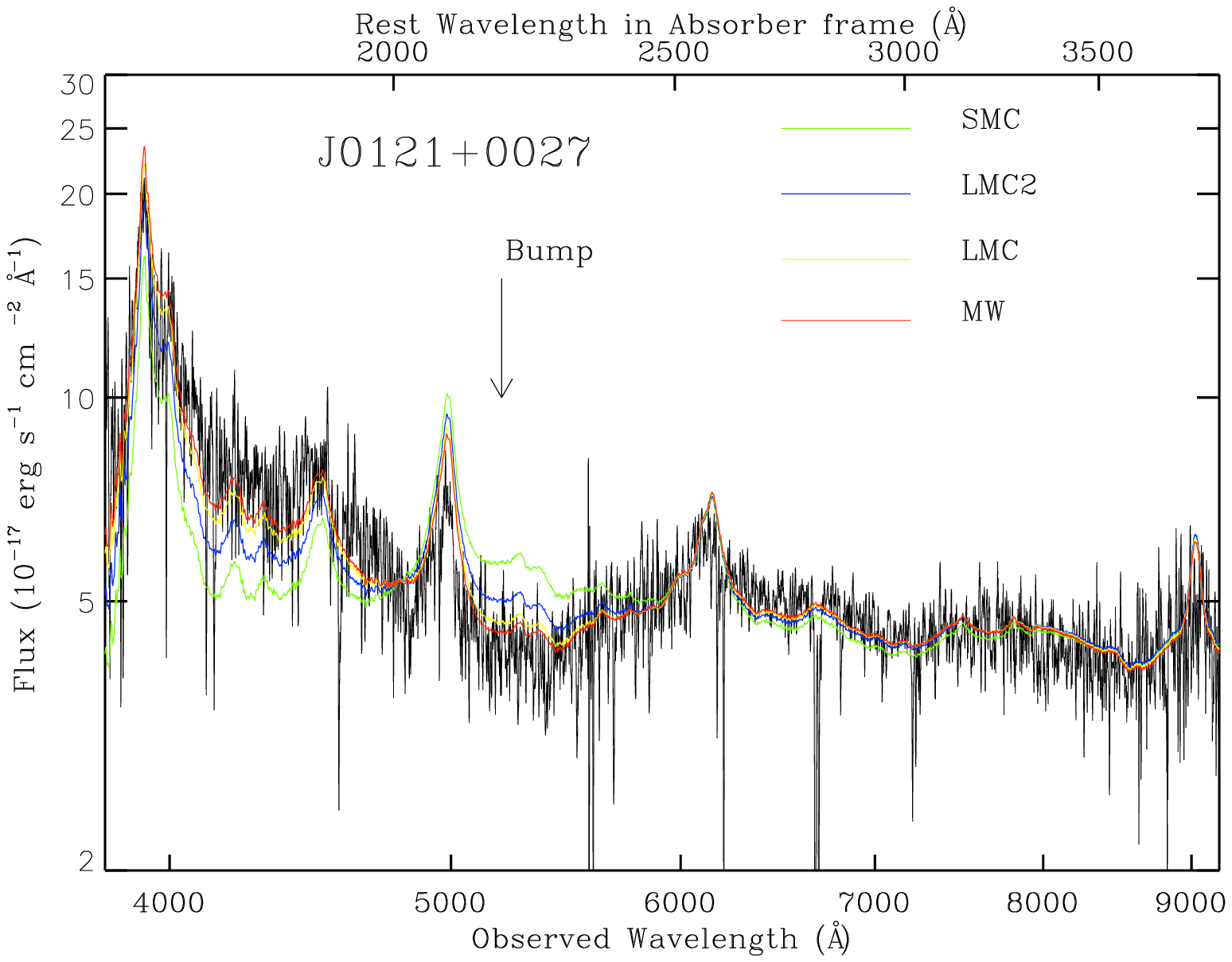}
\caption{Fitting the reddened quasar spectra by reddening the SDSS composite quasar spectrum with different extinction curves. Green spectra are for the average SMC extinction curve; blue spectra are for the average LMC Supershell (LMC2) extinction curve; yellow spectra are for the average LMC extinction curve; red spectra are for the average MW extinction curve. Left panel presents the models for J1459+0024 and right panel presents the models for J0121+0027.\label{fig3}}
\end{figure}

\clearpage

\begin{figure}
\plotone{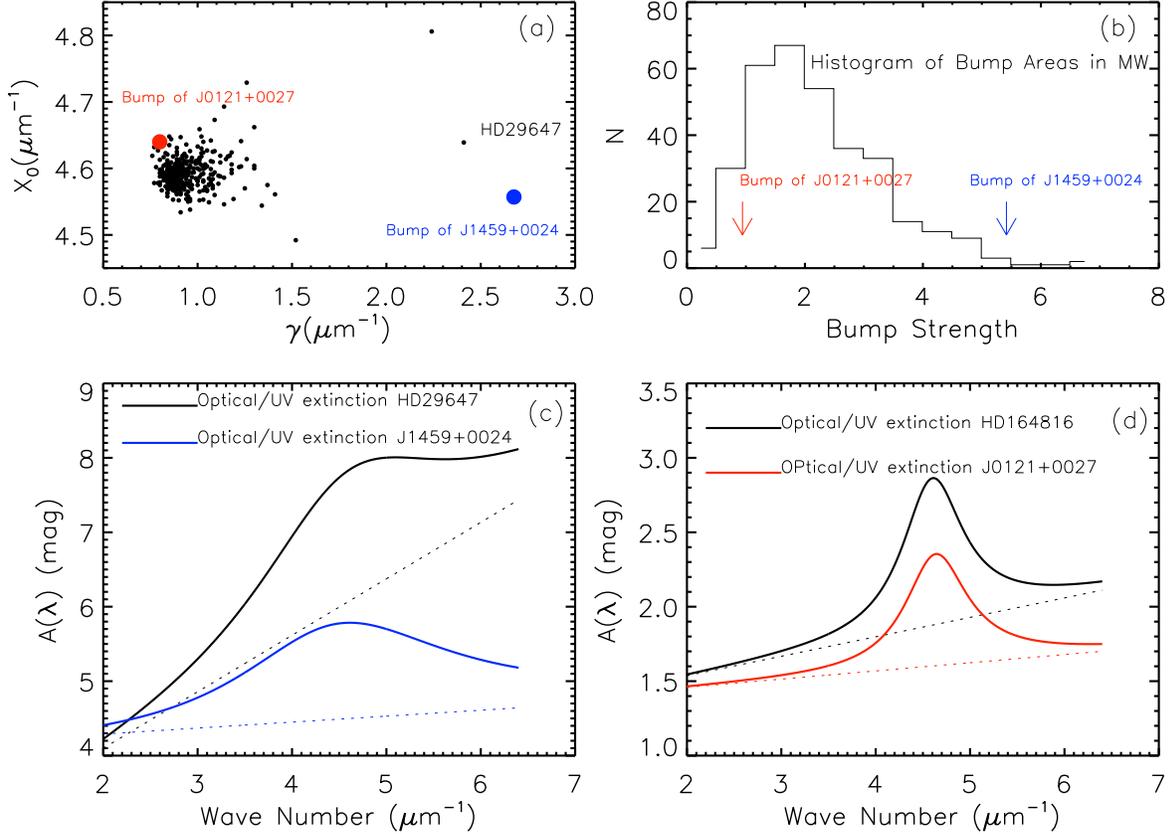}
\caption{Comparisons of extinction bump parameters with the 328 Galactic values measured toward stars from Fitzpatrick \& Massa (2007). (a): the distribution of peak position and width of the bumps. The width of bump in J1459+0024 is broader than the broadest Galactic bump in the line of sight toward star HD29647. (b): Histogram of bump strength measured toward Galactic stars. Red arrow indicates the strength of bump in J0121+0027 and blue arrow indicates the strength of bump in J1459+0024. (c). Direct comparison of derived Optical/UV extinction curve of J1459+0024 with the curve measured in the line of sight toward star HD29647. (d). Direct comparison of derived UV extinction curve of J0121+0027 with the curve measured in the line of sight toward star HD164816 (see the parameters of extinction curves in Table 1). An arbitrary normalization is added to our relative extinction curve when plotting.\label{fig4}}
\end{figure}
\clearpage

\begin{figure}
\plotone{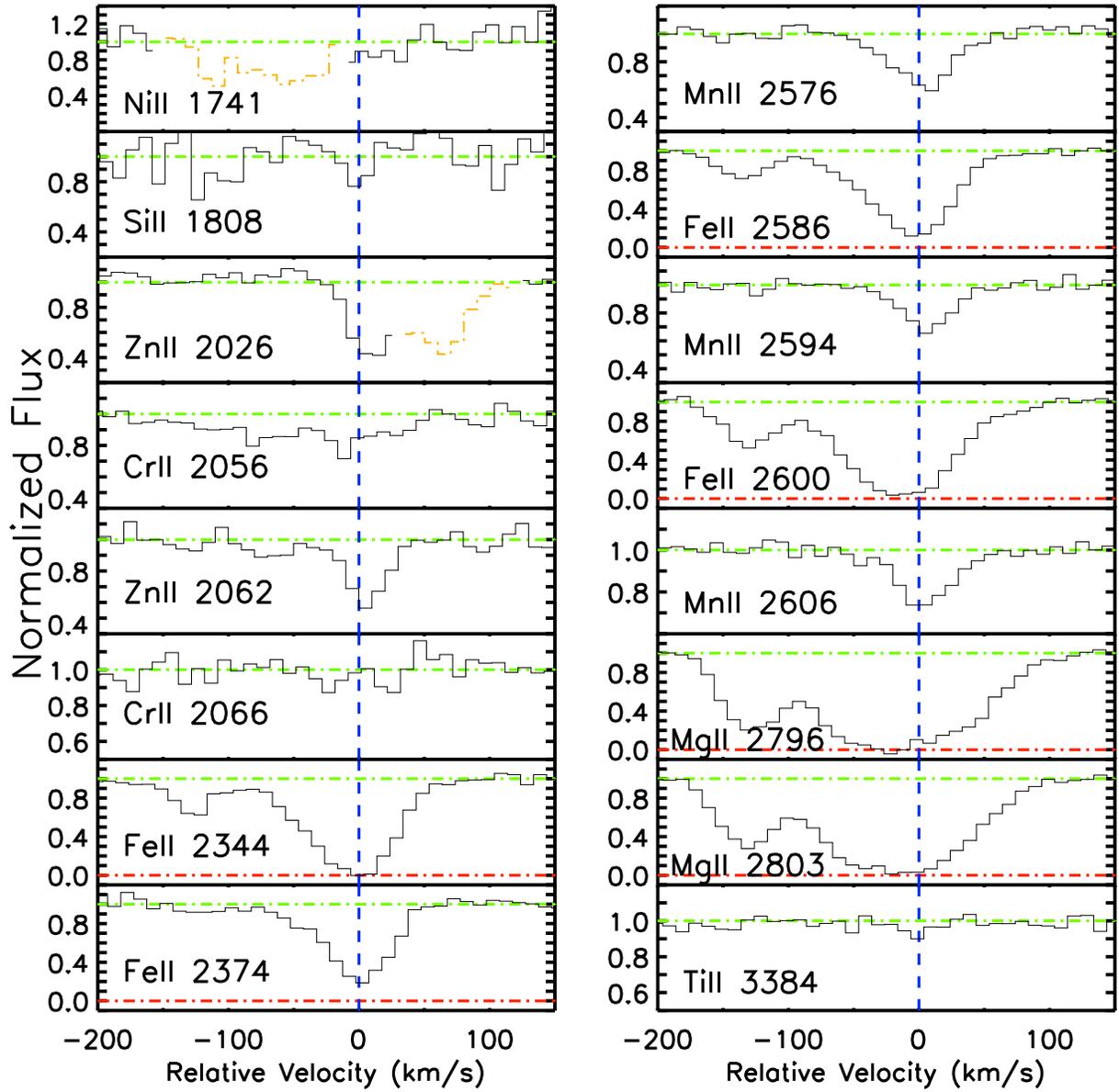}
\caption{Velocity plots of the metal line transitions for absorption system at $z$=1.3947 toward quasar J1459+0024. The vertical thick blue dash lines are corresponding velocity=0 km s$^{-1}$ at that redshift.
\label{fig5}}
\end{figure}
\clearpage

\begin{figure}
\plotone{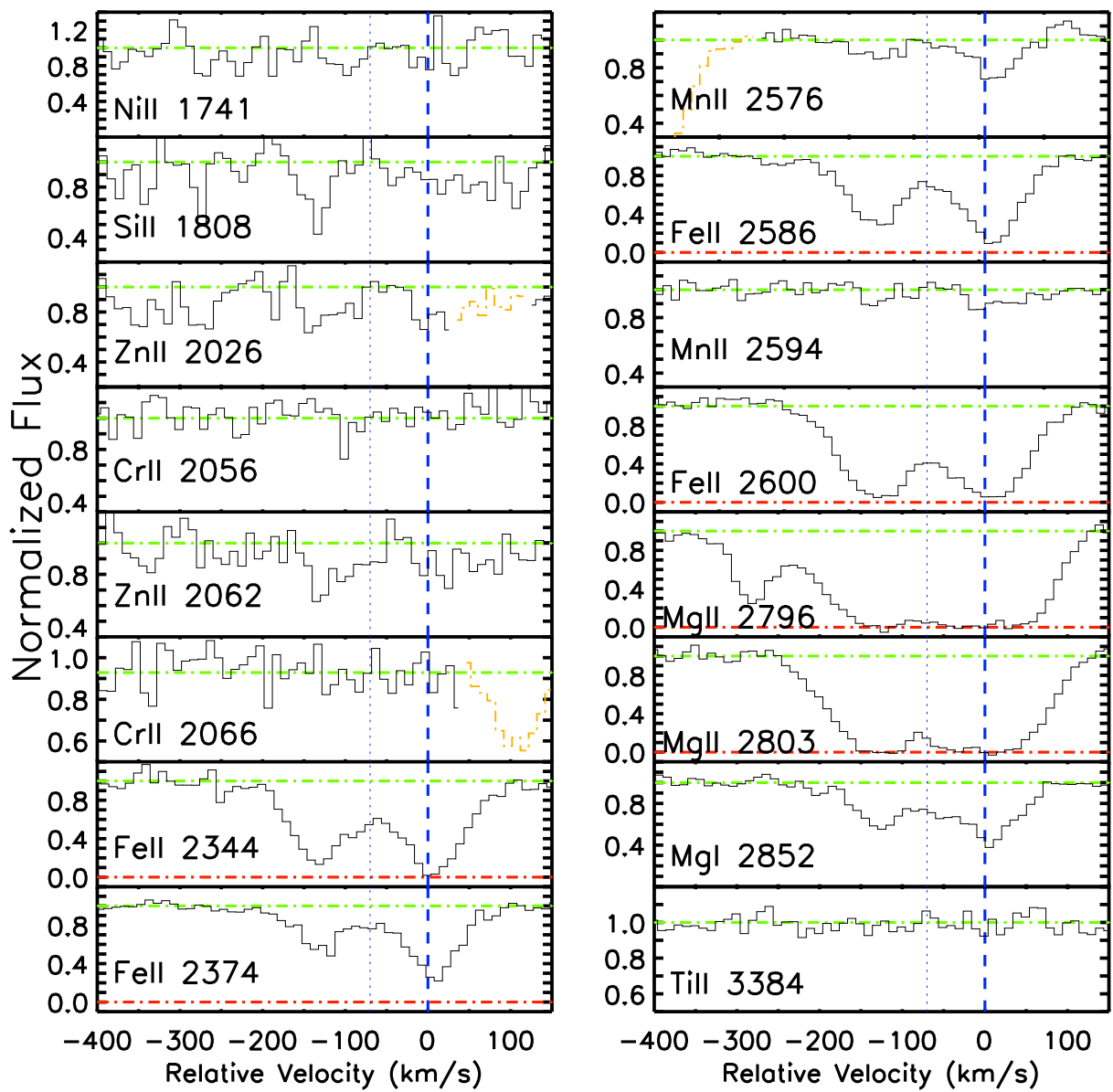}
\caption{Velocity plots of the metal line transitions for absorption system at $z$=1.3888 toward quasar J0121+0027. The vertical thick blue dash lines are corresponding velocity=0 km s$^{-1}$ at that redshift. The thin blue dot lines are corresponding velocity=$-70$ km s$^{-1}$ at $z$=1.3888 to separate the two velocity components.
\label{fig6}}
\end{figure}
\clearpage

\begin{figure}
\plotone{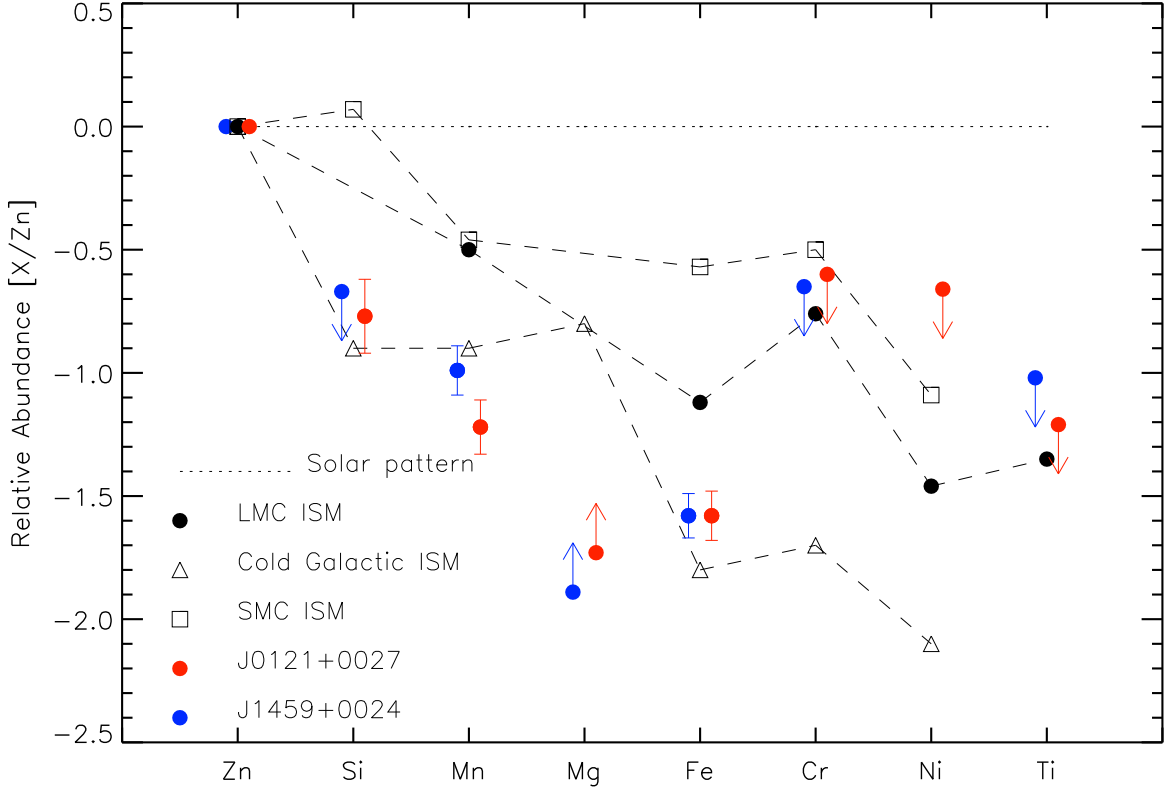}
\caption{Dust depletion patterns in the two 2175-{\AA} absorbers are compared with those measured in ISM of the local group. The blue dots are relative abundance of absorption system toward the quasar J1459+0024. The red dots are relative abundance (over the whole absorption line profile) toward the quasar J0121+0027. The empty squares are dust depletion patterns measured in SMC ISM; the black filled circles are patterns in LMC ISM; the empty triangles are patterns in cold Galactic disk clouds (Welty et al. 1997; Welty et al. 1999).
\label{fig7}}
\end{figure}
\clearpage

\begin{figure}
\plotone{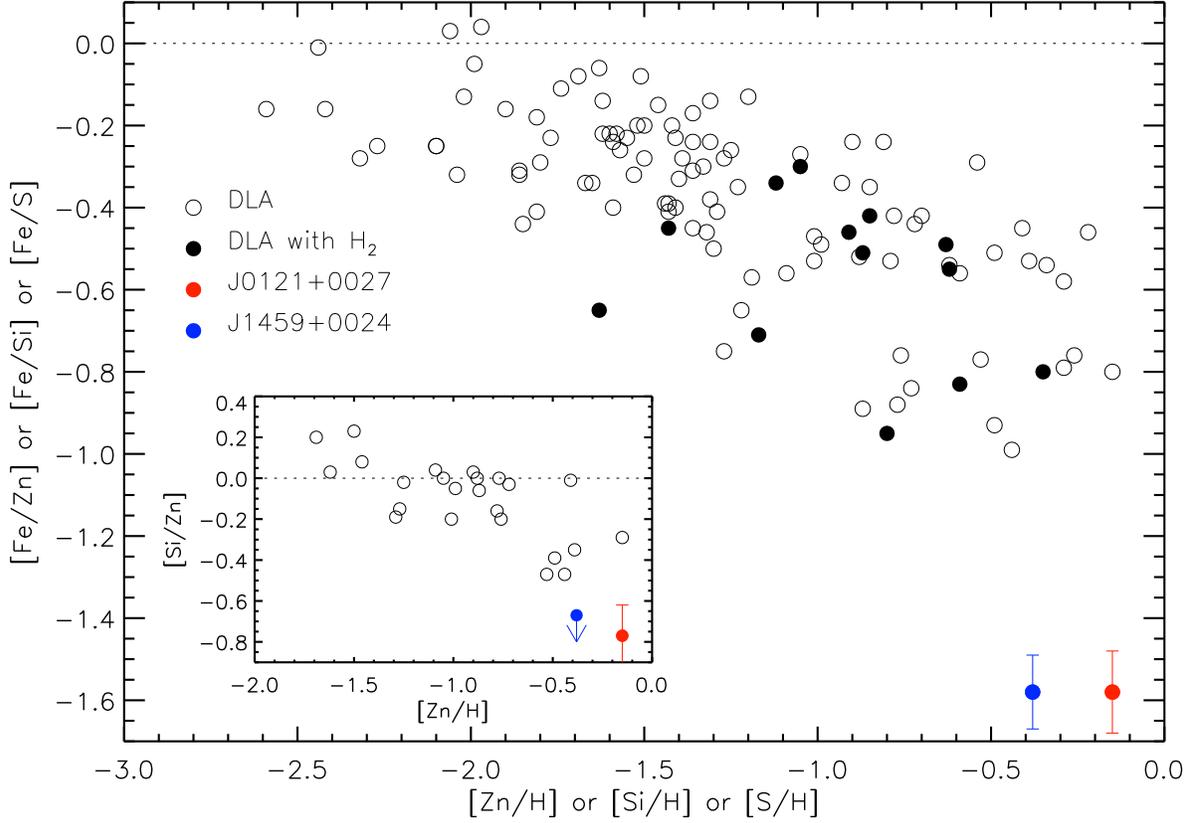}
\caption{Comparison of dust depletion ([Fe/Zn] and [Si/Zn]) in this work with that in a DLA/sub-DLA sample. The empty circles are measurements in a combined DLA/sub-DLA sample having high resolution spectra (Prochaska et al. 2007 and Noterdaeme et al. 2008), while the black filled circles are for H$_2$-bearing DLAs in the subsample from Noterdaeme et al. (2008). N(H~I)=10$^{21}$ cm$^{-1}$ is assumed in both of the 2175-{\AA} absorbers arbitrarily for drawing purpose only.
\label{fig8}}
\end{figure}
\clearpage

\begin{figure}
\plotone{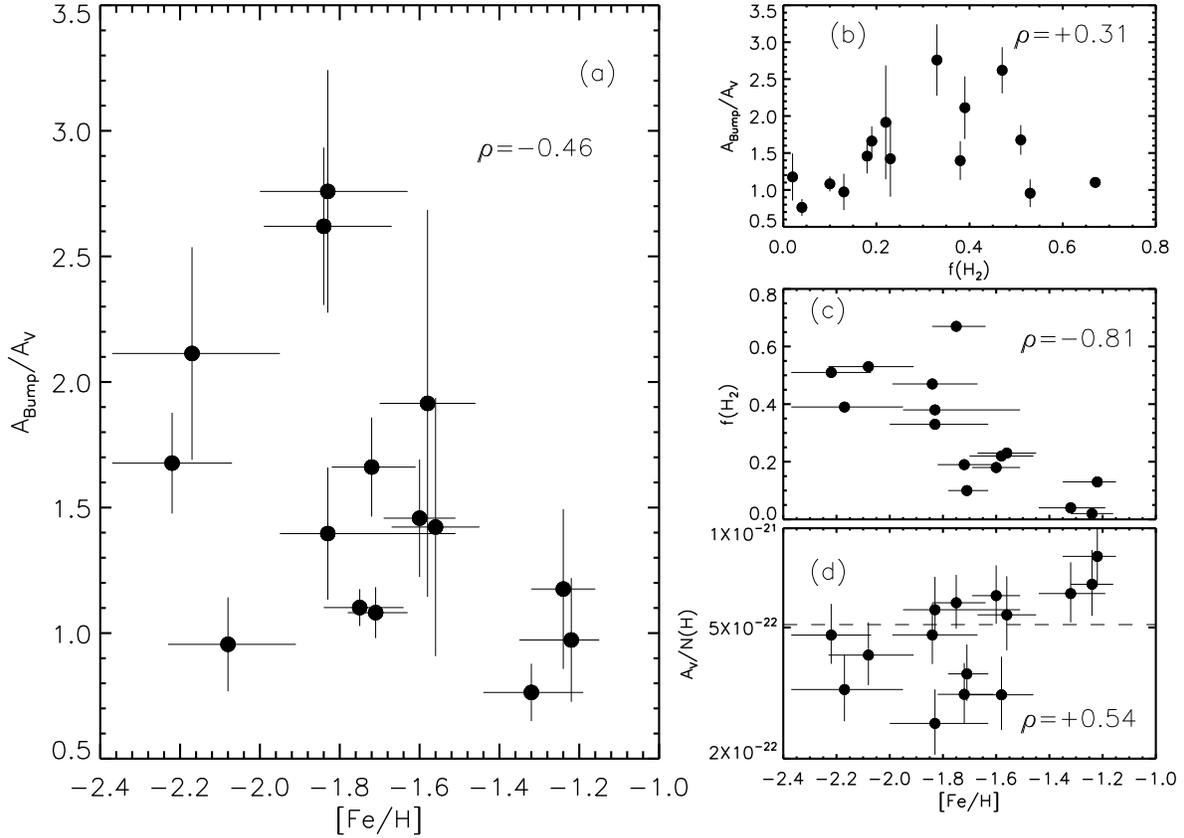}
\caption{Dust depletion and dust extinction features in 15 lines of sight in MW. (a): a tentative anti-correlation between the dust depletions ([Fe/H]) and the relative bump strengths (Spearman's $\rho=-0.46$). (b): a tentative correlation between the fractions of molecular hydrogen (f(H$_2$)) and the relative bump strengths (Spearman's $\rho$=+0.31). (c): a moderate anti-correlation between [Fe/H] and f(H$_2$) in the Galactic clouds (Spearman's $\rho=-0.81$). (d): a tentative correlation between [Fe/H] and dust-to-gas ratio (A$_V$/N(H$_{tot}$); Spearman's $\rho$=0.54). The dash line indicates the average Galactic dust-to-gas ratio (Bohlin et al. 1978). 
\label{fig9}}
\end{figure}
\clearpage

\begin{deluxetable}{lcccccccc}
\tabletypesize{\scriptsize}
\tablecaption{Parameters of Optical/UV Extinction Curves\label{tbl-1}}
\tablewidth{0pt}
\tablehead{\colhead{Reddened Object} & \colhead{$c_1$} & \colhead{$c_2$} & \colhead{$c_3$} & \colhead{$x_0$} & \colhead{$\gamma$} & \colhead{E(\bv)} & \colhead{$R_V$} & \colhead{$\chi_{\nu}^2$}}
\startdata
J1459+0024 & -2.17$\pm$0.02 & 0.08$\pm$0.01 & 9.22$\pm$0.71 & 4.56$\pm$0.02 & 2.68$\pm$0.07 &  &  & 2.19 \\
J0121+0027 & -0.65$\pm$0.02 & 0.06$\pm$0.01 & 0.48$\pm$0.04 & 4.64$\pm$0.01 & 0.80$\pm$0.04 &  &  & 1.08 \\
\tableline
\tableline
HD29647 & -0.77$\pm$0.18 & 0.79$\pm$0.03 & 10.67$\pm$2.79 & 4.639$\pm$0.037 &2.41$\pm$0.22  & 0.96$\pm$0.02 & 3.46$\pm$0.09 & \\
HD164816 & 1.14$\pm$0.39 & 0.45$\pm$0.06 & 2.07$\pm$0.12 & 4.602$\pm$0.004 & 0.78$\pm$0.02 & 0.29$\pm$0.01 & 3.26$\pm$0.18 & \\
\enddata
\tablecomments{Best fitted parameters of extinction curves for the two absorbers. The parameters for reddened Galactic stars are adopted from Fitzpatrick \& Massa (2007). The parameters $c_1$, $c_2$ and $c_3$ for the absorbers are not normalized by E(\bv) and $A_V$, while the relevant parameters for the Galactic stars are normalized values.}
\end{deluxetable}

\begin{deluxetable}{ccccc}
\tabletypesize{\normalsize}
\tablecaption{Results of Fitting SDSS Spectra with Different Extinction Curves\label{tbl-2}}
\tablewidth{0pt}
\tablehead{
\colhead{Extinction Law} & \colhead{E(\bv)} & \colhead{E(\bv)} & \colhead{$\chi_{\nu}^2$} & \colhead{$\chi_{\nu}^2$} \\
\colhead{} & \colhead{(J1459)} & \colhead{(J0121)} & \colhead{(J1459)} & \colhead{(J0121)}
}
\startdata
SMC & 0.21$\pm$0.01 & 0.11$\pm$0.01 & 6.34 & 1.69 \\
LMC2 & 0.29$\pm$0.01 & 0.17$\pm$0.01 & 4.26 & 1.28 \\
LMC & 0.25$\pm$0.01 & 0.14$\pm$0.01 & 3.88 & 1.14 \\
MW & 0.26$\pm$0.01 & 0.15$\pm$0.1 & 3.93 & 1.10 \\
\enddata
\tablecomments{Results of fitting the Spectra by reddening the SDSS composite quasar spectrum with the average extinction curves of SMC Bar, LMC Supershell (LMC2), LMC and MW. The average extinction curves of SMC, LMC2 and LMC are adopted from Gordon et al. 2003; the average extinction curve of MW is adopted from Fitzpatrick \& Massa 2007.}
\end{deluxetable}

\begin{deluxetable}{llccccc}
\tabletypesize{\scriptsize}
\tablecaption{Measurements of Column Densities\label{tbl-3}}
\tablewidth{0pt}
\tablehead{
\colhead{$\lambda_{vacuum}$} & \colhead{Ion} & \colhead{$f$} & \colhead{N$_X^a$} & \colhead{N$_X^b$} & \colhead{N$_X^c$} & \colhead{N$_X^d$} \\
\colhead{(\AA)} & \colhead{} & \colhead{} & \colhead{log(cm$^{-2}$)} & \colhead{log(cm$^{-2}$)} & \colhead{log(cm$^{-2}$)} & \colhead{log(cm$^{-2}$)}
}
\startdata
1741.5531 & Ni \scriptsize{\uppercase\expandafter{\romannumeral2}} & 0.0427 & \ldots$^e$ & $<$14.42 & $<$14.30 & $<$14.13 \\
1808.0130 & Si \scriptsize{\uppercase\expandafter{\romannumeral2}} & 0.0022 & $<$15.46 & 15.59$\pm$0.13 & 15.22$\pm$0.21 & 15.35$\pm$0.17 \\
1854.7164 & Al \scriptsize{\uppercase\expandafter{\romannumeral3}} & 0.5390 & $<$13.81 & 13.71$\pm$0.06 & 13.03$\pm$0.09 & 13.61$\pm$0.06 \\
1862.7895 & Al \scriptsize{\uppercase\expandafter{\romannumeral3}} & 0.2680 & $<$13.49 & 13.55$\pm$0.09 & 12.94$\pm$0.19 & 13.43$\pm$0.09 \\
2026.1360 & Zn \scriptsize{\uppercase\expandafter{\romannumeral2}} & 0.4890 & 13.12$\pm$0.06 & 13.26$\pm$0.08 & 12.94$\pm$0.10 & 12.98$\pm$0.10\\
2056.2539 & Cr \scriptsize{\uppercase\expandafter{\romannumeral2}} & 0.1050 & $<$13.81 & $<$13.33 & $<$13.28 & $<$13.20 \\
2062.6640 & Zn \scriptsize{\uppercase\expandafter{\romannumeral2}} & 0.2560 & 13.22$\pm$0.08 & 13.45$\pm$0.09 & 13.19$\pm$0.11 & 13.12$\pm$0.13 \\
2066.1610 & Cr \scriptsize{\uppercase\expandafter{\romannumeral2}} & 0.0515 & $<$13.61 & $<$13.89 & $<$13.73 & $<$13.68\\
2344.2140 & Fe \scriptsize{\uppercase\expandafter{\romannumeral2}} & 0.1140 & $>$14.50 & $>$14.69 & $>$14.29 & $>$14.48\\
2374.4612 & Fe \scriptsize{\uppercase\expandafter{\romannumeral2}} & 0.0313 & 14.61$\pm$0.05 & 14.82$\pm$0.05 & 14.36$\pm$0.06 & 14.64$\pm$0.06 \\
2382.7650 & Fe \scriptsize{\uppercase\expandafter{\romannumeral2}} & 0.3200 & $>$14.15 & $>$14.41 & $>$14.09 & $>$14.13 \\
2576.8770 & Mn \scriptsize{\uppercase\expandafter{\romannumeral2}} & 0.3508 & 13.02$\pm$0.06 & 13.07$\pm$0.06 & 12.55$\pm$0.10 & 12.92$\pm$0.07\\
2586.6500 & Fe \scriptsize{\uppercase\expandafter{\romannumeral2}} & 0.0691 & 14.46$\pm$0.05 & 14.70$\pm$0.05 & 14.31$\pm$0.05 & 14.47$\pm$0.05\\
2594.4990 & Mn \scriptsize{\uppercase\expandafter{\romannumeral2}} & 0.2710 & 13.00$\pm$0.07 & 12.80$\pm$0.11 & 12.11$\pm$0.42 & 12.70$\pm$0.11\\
2600.1729 & Fe \scriptsize{\uppercase\expandafter{\romannumeral2}} & 0.2390 & $>$14.25 & $>$14.44 & $>$14.14 & $>$14.13 \\
2606.4620 & Mn \scriptsize{\uppercase\expandafter{\romannumeral2}} & 0.1927 & 13.07$\pm$0.08 & 12.83$\pm$0.14 & 12.44$\pm$0.26 & 12.60$\pm$0.18\\
2796.3520 & Mg \scriptsize{\uppercase\expandafter{\romannumeral2}} & 0.6123 & $>$14.03 & $>$14.42 & $>$14.09 & $>$14.14 \\
2803.5310 & Mg \scriptsize{\uppercase\expandafter{\romannumeral2}} & 0.3054 & $>$14.26 & $>$14.65 & $>$14.25 & $>$14.43\\
2852.9642 & Mg \scriptsize{\uppercase\expandafter{\romannumeral1}} & 1.8100 & 12.67$\pm$0.05 & 12.91$\pm$0.05 & 12.53$\pm$0.05 & 12.68$\pm$0.05\\
3384.7400 & Ti \scriptsize{\uppercase\expandafter{\romannumeral2}} & 0.3580 & $<$12.50 & $<$12.54 & $<$12.35 & $<$12.36 \\
\enddata
\tablecomments{Vacuum wavelengths and oscillator strength $f$ are adopted from the Atomic Data conducted by J. X. Prochaska (http://www.astro.ufl.edu/~jpaty/qal.lst). The systematic error of column densities can exceed 0.05 dex due to continuum fitting and line saturation with ESI data. We combine the photon noise error and the estimated systematic error quadratically as reported error.}
\tablenotetext{a}{Absorption lines system at $z$=1.3947 toward the QSO J1459+0024}
\tablenotetext{b}{Absorption lines system at $z$=1.3888 toward the QSO J0121+0027}
\tablenotetext{c}{The blue component in profiles of absorption lines toward the QSO J0121+0027}
\tablenotetext{d}{The red component in profiles of absorption lines toward the QSO J0121+0027}
\tablenotetext{e}{Ni~II 1741 absorption line in J1459+0024 system is blended with an unidentified absorption feature (see Figure 5). Its column density cannot be measured.}
\end{deluxetable}

\begin{deluxetable}{cccccccccc}
\tabletypesize{\scriptsize}
\tablecaption{Measurements of Dust Depletion Levels\label{tbl-4}}
\tablewidth{0pt}
\tablehead{
\colhead{} & \colhead{J1459} & \colhead{J0121$^a$} & \colhead{J0121$^b$} & \colhead{J0121$^c$} & \colhead{Cold$^d$} & \colhead{Warm$^d$} & \colhead{Halo$^d$} & \colhead{LMC$^e$} & \colhead{SMC$^f$}\\
\colhead{Species} & \colhead{} & \colhead{} & \colhead{} & \colhead{[X/Zn]} &
\colhead{} & \colhead{} & \colhead{} & \colhead{} & \colhead{}
}
\startdata
Si & $<$-0.67 & -0.77$\pm$0.16 & -0.88$\pm$0.24 & -0.68$\pm$0.21 & -0.9 & -0.2 & -0.2 & $<$+0.32 & +0.07 \\ 
Mn & -0.99$\pm$0.10 & -1.22$\pm$0.11 & -1.46$\pm$0.12 & -1.05$\pm$0.15 & -1.0 & -0.7 & -0.5 & -0.50 & -0.46 \\
Mg & $>$-1.89 & $>$-1.73 & $>$-1.87 & $>$-1.62 & -0.8 & -0.4 & \ldots & \ldots & \ldots \\
Fe & -1.58$\pm$0.09 & -1.58$\pm$0.10 & -1.72$\pm$0.12 & -1.45$\pm$0.14 & -1.8 & -1.2 & -0.5 & -1.12 & -0.57 \\
Cr & $<$-0.65 & $<$-0.60 &$<$-0.50 & $<$-0.48 & -1.7 & -1.0 & -0.5 & -0.76 & -0.50 \\
Ni & \ldots & $<$-0.66 & $<$-0.52 & $<$-0.62 & -2.1 & -1.5 & -0.7 & -1.46 & -1.09 \\
Ti & $<$-1.02 & $<$-1.21 & $<$-1.14 & $<$-1.06 & -2.5 & -1.1 & -0.7 &  -1.35 & \ldots \\
\enddata
\tablecomments{The solar photospheric values are adopted from Asplund et al. (2005).}
\tablenotetext{a}{Depletion in overall profiles of absorption lines toward the QSO J0121+0027}
\tablenotetext{b}{Depletion in the blue component in profiles of absorption lines toward the QSO J0121+0027}
\tablenotetext{c}{Depletion in the red component in profiles of absorption lines toward the QSO J0121+0027}
\tablenotetext{d}{Depletion in cold disk clouds, warm disk clouds, diffuse halo clouds (Jenkins et al. 1986; Welty et al. 1999 and Welty et al. 2001)}
\tablenotetext{e}{Depletion in LMC ISM in line of sight toward SN 1987A (Welty et al. 1999)}
\tablenotetext{f}{Depletion in SMC ISM in line of sight toward star SK108 (Welty et al. 1997)}
\end{deluxetable}

\begin{deluxetable}{lcccccccc}
\tabletypesize{\scriptsize}
\tablecaption{Fe~II Abundance and Extinction Parameters in MW\label{tbl-5}}
\tablewidth{0pt}
\tablehead{
\colhead{Star} & \colhead{N(H$_{tot}$)$^a$} & \colhead{f(H$_2$)} & \colhead{N(FeII)$^b$} &
\colhead{R$_V$} & \colhead{E(\bv)} & \colhead{c$_3$} & \colhead{$\gamma$} & \colhead{Reference} \\
\colhead{} & \colhead{log(cm$^{-2}$)} & \colhead{} & \colhead{log(cm$^{-2}$)} & \colhead{} &
\colhead{mag} & \colhead{} & \colhead{$\mu$m$^{-1}$} & \colhead{} \\
\colhead{(1)} & \colhead{(2)} & \colhead{(3)} & \colhead{(4)} & \colhead{(5)} &
\colhead{(6)} & \colhead{(7)} & \colhead{(8)} & \colhead{(9)}
}
\startdata
HD12323 & 21.29 & 0.22 & 15.16$^{+0.07}_{-0.07}$ & 2.90$\pm$0.26& 0.21$\pm$0.01 & 3.04$\pm$0.24& 0.86$\pm$0.02 & 2,3,4\\
HD37903 & 21.50 & 0.53 & 14.87$^{+0.12}_{-0.10}$ & 3.95$\pm$0.21& 0.33$\pm$0.01 & 2.21$\pm$0.13& 0.92$\pm$0.02 & 2,3,4\\
HD40893 & 21.59 & 0.19 & 15.32$^{+0.06}_{-0.05}$ & 2.71$\pm$0.13& 0.45$\pm$0.01 & 2.35$\pm$0.08& 0.82$\pm$0.01 & 2,3,4\\
HD73882 & 21.59 & 0.67 & 15.29$^{+0.06}_{-0.04}$ & 3.45$\pm$0.08& 0.67$\pm$0.01 & 2.71$\pm$0.10& 1.12$\pm$0.02 & 2,3,4\\
HD91651 & 21.16 & 0.02 & 15.37$^{+0.03}_{-0.03}$ & 3.49$\pm$0.22& 0.28$\pm$0.01 & 2.69$\pm$0.26& 1.03$\pm$0.05 & 2,3,4\\
HD93222 & 21.42 & 0.04 & 15.55$^{+0.08}_{-0.07}$ & 5.05$\pm$0.20& 0.33$\pm$0.01 & 1.94$\pm$0.10& 0.79$\pm$0.02 & 2,3,4\\
HD104705 & 21.17 & 0.13 & 15.40$^{+0.02}_{-0.08}$ & 4.34$\pm$0.20& 0.28$\pm$0.01 & 2.15$\pm$0.13& 0.80$\pm$0.01 & 2,3,4\\
HD147888 & 21.76 & 0.10 & 15.50$^{+0.03}_{-0.02}$ & 4.08$\pm$0.13& 0.51$\pm$0.01 & 2.50$\pm$0.11& 0.89$\pm$0.02 & 2,3,4\\
HD170740 & 21.46 & 0.51 & 14.69$^{+0.10}_{-0.10}$ & 2.91$\pm$0.14& 0.47$\pm$0.01 & 2.92$\pm$0.10& 0.94$\pm$0.01 & 1,2,4\\
HD177989 & 21.06 & 0.23 & 14.95$^{+0.06}_{-0.06}$ & 2.85$\pm$0.27& 0.22$\pm$0.01 & 2.40$\pm$0.22& 0.93$\pm$0.04 & 2,3,4\\
HD179406 & 21.44 & 0.39 & 14.72$^{+0.17}_{-0.15}$ & 2.88$\pm$0.20& 0.31$\pm$0.01 & 3.72$\pm$0.23& 0.96$\pm$0.03 & 2,3,4\\
HD185418 & 21.39 & 0.47 & 15.00$^{+0.12}_{-0.10}$ & 2.48$\pm$0.13& 0.47$\pm$0.01 & 3.64$\pm$0.12& 0.88$\pm$0.01 & 1,2,4\\
HD197512 & 21.44 & 0.33 & 15.06$^{+0.15}_{-0.12}$ & 2.43$\pm$0.15& 0.29$\pm$0.01 & 4.14$\pm$0.18& 0.97$\pm$0.02 & 1,2,4\\
HD199579 & 21.25 & 0.38 & 14.87$^{+0.27}_{-0.07}$ & 3.05$\pm$0.17& 0.33$\pm$0.01 & 2.63$\pm$0.17& 0.97$\pm$0.03 & 1,2,4\\
HD209339 & 21.24 & 0.18 & 15.09$^{+0.04}_{-0.04}$ & 3.10$\pm$0.10& 0.35$\pm$0.01 & 2.33$\pm$0.11& 0.81$\pm$0.01 & 2,3,4\\
\enddata
\tablecomments{Column (1) is the HD name of background early type Galactic stars. Column (2), (3) and (4) are adopted from reference 1, 2 and 4. Column (5), (6), (7) and (8) are adopted from reference 3.}
\tablenotetext{a}{N(H$_{tot}$)=N(HI)+N(H$_2$) Column densities are measured by Voigt profile fitting.}
\tablenotetext{b}{Initial N(FeII) was measured by curve of growth analysis of FeII weak/moderate strength transitions in Snow et al. (2002) and Jensen \& Snow (2007). Jenkins (2009) corrected the N(FeII) by using most modern FeII $f$-values. We adopt the corrected N(FeII) in this work.}
\tablerefs{(1) Snow et al. 2002; (2) Fitzpatrick \& Massa 2007; (3) Jensen \& Snow 2007; (4) Jenkins 2009}
\end{deluxetable}
\end{document}